\documentclass[letterpaper, 10 pt, conference]{ieeeconf}

\IEEEoverridecommandlockouts
\usepackage[utf8]{inputenc}
\usepackage[T1]{fontenc}
\usepackage{amsmath}
\usepackage{booktabs}
\usepackage{graphicx}
\usepackage{cite}
\usepackage{textcomp}
\usepackage{newtxtext,newtxmath}
\usepackage{fancyhdr}

\makeatletter
\@twosidetrue
\makeatother

\fancypagestyle{RALheader}{%
    \fancyhf{}

    \fancyhead[L]{%
        \footnotesize %
    }

    \fancyhead[R]{%
        \footnotesize\thepage%
    }

    \setlength{\headheight}{14pt}
    \setlength{\headsep}{6pt}
}

\fancypagestyle{RALoddEven}{%
    \fancyhf{}

    \fancyhead[LE]{%
        \footnotesize IEEE ROBOTICS AND AUTOMATION LETTERS. PREPRINT VERSION%
    }

    \fancyhead[RE]{%
        \footnotesize\thepage%
    }

    \fancyhead[LO]{%
        \footnotesize\thepage%
    }

    \fancyhead[RO]{%
        \footnotesize Jin et al.: Underwater MMG-Based Muscle State Monitoring with Integrated Emergency Buoyancy Assistance%
    }

    \setlength{\headheight}{14pt}
    \setlength{\headsep}{6pt}
}

\pdfmapfile{=pdftex.map}

\title{\LARGE \bf
Underwater MMG-Based Muscle State Monitoring with Integrated Emergency Buoyancy Assistance
}

\author{
Xiao Jin$^{1}$,
Yixian Fan$^{2,3}$,
Zefeng Yuan$^{2,3}$,
and Zhenhua Yu$^{1,2,3}$%
\thanks{%
This research is partially supported by the Capital Equipment funding of
the University of Aberdeen CF10720-10.
(\textit{Corresponding author: Zhenhua Yu, zhenhua.yu@abdn.ac.uk}).
}%
\thanks{%
$^{1}$Dyson School of Design Engineering,
Imperial College London,
London SW7 2AZ, UK.
}%
\thanks{%
$^{2}$Aberdeen Institute of Data Science and Artificial Intelligence,
South China Normal University,
Foshan, China.
}%
\thanks{%
$^{3}$Department of Computer Science,
University of Aberdeen,
Aberdeen AB24 3UE, UK.
}%
}

\begin{document}

\maketitle

\begin{abstract}
This paper presents an underwater MMG-driven wearable emergency assistance system for lower-leg muscle-state monitoring and automatic buoyancy deployment. A compact microphone-based MMG sensor was waterproofed using a flexible 5 mil PE membrane, preserving identifiable muscle-vibration responses under immersion, depth variation, and stirring disturbances. Two lower-leg sensors captured stroke-dependent MMG patterns across four swimming styles, and a MiniRocket classifier achieved 91.91\% window-level and 97.56\% file-level accuracy. For cramp-related monitoring, a pattern-based risk score was used to identify representative pre-cramp abnormal muscle-state transitions during rhythmic motion. A controlled underwater test demonstrated the closed sensing--decision--actuation chain, triggering CO$_2$ release, airbag inflation, and flotation in less than 5~s. These results support underwater MMG as a sensing basis for wearable robotic emergency assistance in aquatic environments.

\end{abstract}

\section{INTRODUCTION}
Water-related accidents remain a persistent safety concern even in highly controlled environments such as swimming pools, where incidents including muscle cramps, involuntary inhalation, and sudden loss of motor control can rapidly escalate into life-threatening situations. Globally, drowning accounts for hundreds of thousands of fatalities each year\cite{
Kao2024Next-Generation,Jalalifar2022A}, many of which occur without timely detection or intervention\cite{Jalalifar2024Enhancing}, highlighting the limitations of current supervision- and vision-based monitoring approaches\cite{Shatnawi2024Advances}. In response, substantial efforts and investments have been directed toward water safety and rescue technologies, spanning lifeguard augmentation systems, automated detection platforms, and intelligent rescue devices. Despite this progress, reliably identifying early-stage physiological distress under water—prior to visible failure or loss of buoyancy—remains an open challenge, motivating the development of sensing and intervention mechanisms that can operate directly at the human–water interface\cite{Zhang2025A,Zhu2025A}.

In most public and natural aquatic environments, prevention of water-related accidents still relies primarily on on-site lifeguards and professional rescue teams performing visual monitoring and post-event intervention\cite{Ramos2021The}. Rescue effectiveness follows a reactive chain of abnormality detection, victim approach, and resuscitation, and is highly dependent on observation quality and response time\cite{Bierens2023A}. However, drowning often progresses silently within tens of second\cite{Jalalifar2022A}, while lifeguards predominantly rely on visual or auditory cues, creating an inherent temporal bottleneck. In practice, visibility degradation caused by surface reflections, occlusion, crowd density, and complex water geometries, together with the need to monitor multiple targets simultaneously, further constrains timely detection\cite{Burke2019Requirements}. As a result, manual rescue is widely regarded as necessary but insufficient as a standalone safety mechanism\cite{Ramos2021The} [14,20].

To address these limitations, environment-mounted monitoring systems have been introduced, including camera-based computer vision platforms, water-surface or underwater sensor arrays, and area-level alarm systems\cite{Poongodi2025Hybrid,Lyu2023Unmanned}. Such approaches can reduce detection latency under favorable conditions and extend spatial coverage through fixed or aerial sensing platforms\cite{Jalalifar2022A}. Nevertheless, their performance is strongly influenced by environmental factors such as lighting, turbidity, surface disturbance, and occlusion, and they typically require extensive infrastructure deployment\cite{Yuan2022Marine}. More importantly, these systems primarily detect already externalized abnormal behaviors or loss of buoyancy, and remain limited in identifying early-stage physiological distress—such as muscle fatigue or transient loss of motor control—before visible failure occurs\cite{Monoli2022Wearable,Yang2024An}.

Motivated by the limitations of environment-mounted monitoring, research has increasingly shifted toward on-body wearable systems for continuous, user-centric underwater safety monitoring \cite{Basthikodi2024Automated}. Within wearable sensing, muscle activity provides a direct and interpretable signal for motion intention, force generation, and fatigue \cite{Li2024Non-invasive}. Existing wearable muscle-sensing approaches mainly include sEMG, FMG, and EIT \cite{Zheng2022A}, and muscle signals have been shown to reflect movement and fatigue before observable joint motion \cite{Suo2024AI}. In aquatic studies, sEMG remains the dominant method for analyzing swimming muscle activation \cite{Kwok2022Underwater}. These findings support wearable muscle-based monitoring for underwater safety, but also reveal the limitations of current sensing strategies.

Despite its wide use, EMG is highly sensitive to electrode contact and mechanical disturbances, which limits its reliability in underwater and long-term wearable settings. Changes in skin impedance, sweating, and electrode displacement can degrade signal quality and classification robustness \cite{Woodward2019Segmenting}. In aquatic environments, these issues are further amplified by water-induced electrical shorting and motion artefacts, while existing underwater EMG studies are often limited to short-duration tests and restricted muscle groups \cite{Kwok2022Underwater}. Although encapsulation can partially reduce these problems, it does not eliminate the vulnerability of bioelectrical sensing in wet conditions. This motivates the use of non-electrical muscle sensing based on mechanical vibration or deformation, providing a clear basis for underwater MMG sensing.

In contrast to bioelectrical sensing, mechanomyography (MMG) measures the low-frequency mechanical vibrations produced by muscle contraction, making it less sensitive to electrical shorting, skin--electrode impedance variation, and electromagnetic interference. This property is particularly relevant for underwater wearable monitoring, where stable electrical contact is difficult to maintain and water-induced leakage can degrade bioelectrical signals. Compared with IMU- or pressure-based sensing, which mainly captures limb motion, external load, or hydrodynamic interaction, MMG remains more directly related to the underlying muscle state while avoiding the dependence on skin-surface electrical conduction. Combined with compact microphone-based sensing, MMG therefore provides a mechanically grounded and practical route for underwater muscle-state monitoring.

Building on this basis, this work presents an underwater MMG-driven wearable emergency assistance system that integrates waterproof lower-leg MMG sensing, swimming-motion context recognition, pattern-based cramp-risk scoring, and state-machine-triggered CO$_2$ airbag actuation. The system is designed to identify persistent pre-cramp abnormal muscle-state transitions during rhythmic underwater motion and trigger buoyancy assistance before cramp fully disrupts voluntary movement. These components form a closed sensing--decision--actuation framework for wearable robotic emergency assistance in aquatic environments.

\section{Underwater MMG Sensor Design}
\label{introduction}

\subsection{Structural Design and Waterproof Material Selection}

The proposed sensor was modified from our previous MMG module to improve compactness and underwater robustness. The total height was reduced to approximately 8~mm while maintaining the original diameter, resulting in a low-profile wearable configuration. A 7~mm-diameter and 5~mm-high conical cavity was introduced to enhance vibration coupling and reduce internal acoustic interference~\cite{11399890}. Dragon Skin 30 silicone was filled in the rear cavity to attenuate structural noise~\cite{11276866}. The sensing element is a MEMS capacitive microphone (Knowles SPU1410LR5H-QB), whose frequency response covers the typical MMG range of 10--150~Hz.

For underwater applications, the waterproof layer must provide sufficient sealing while preserving mechanical vibration transmission. Therefore, a flexible polymer membrane was adopted as the acoustic interface. Five candidate membranes with different materials and thicknesses were experimentally compared, including PTFE-based films, polyethylene (PE) films, and aluminized polyester films. Each membrane was attached to the acoustic port, and MMG signals were recorded during finger and wrist extension at 600~Hz sampling frequency.

\begin{figure}[!htbp]
    \centering
    \includegraphics[width=0.9\linewidth]{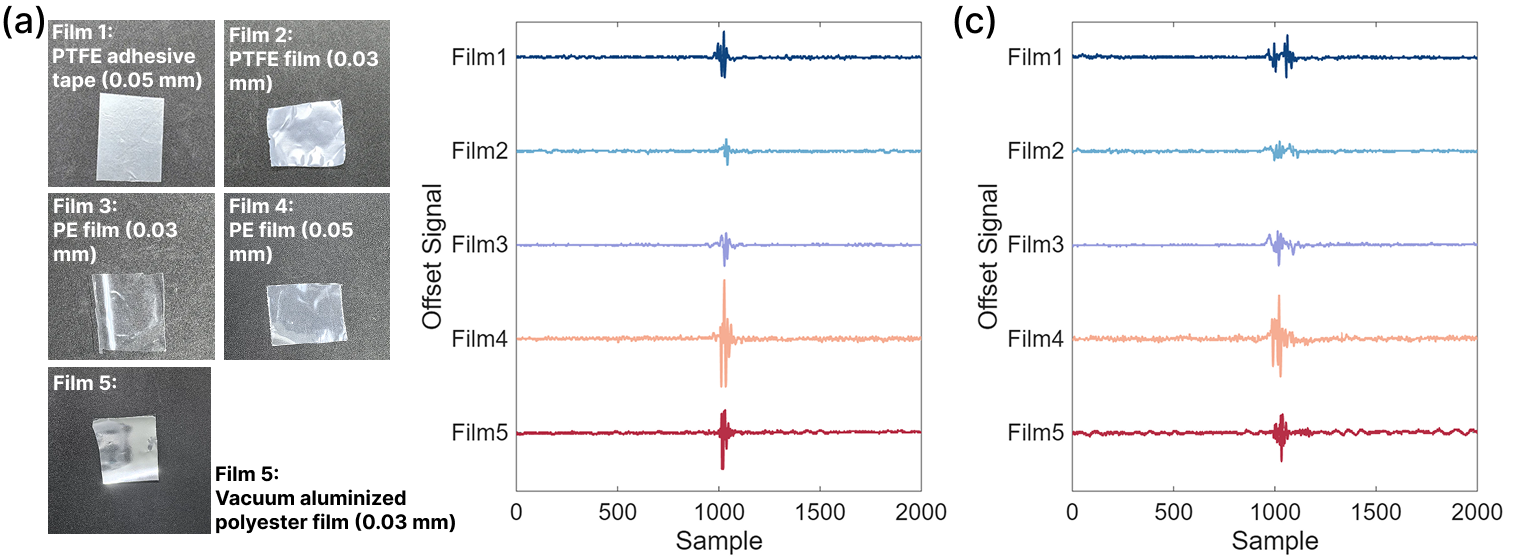}
    \caption{Representative MMG signals measured with different waterproof membranes during finger extension and wrist extension.}
    \label{raw_membrane}
\end{figure}

The measured signals showed that membrane selection significantly influenced vibration transmission. PTFE-based membranes exhibited stronger attenuation, whereas PE membranes provided larger vibration amplitudes and clearer transient responses. Signal features including RMS, MPF, bandwidth, SNR, and peak-to-peak amplitude were further extracted for quantitative comparison.

\begin{figure}[!htbp]
    \centering
    \includegraphics[width=0.9\linewidth]{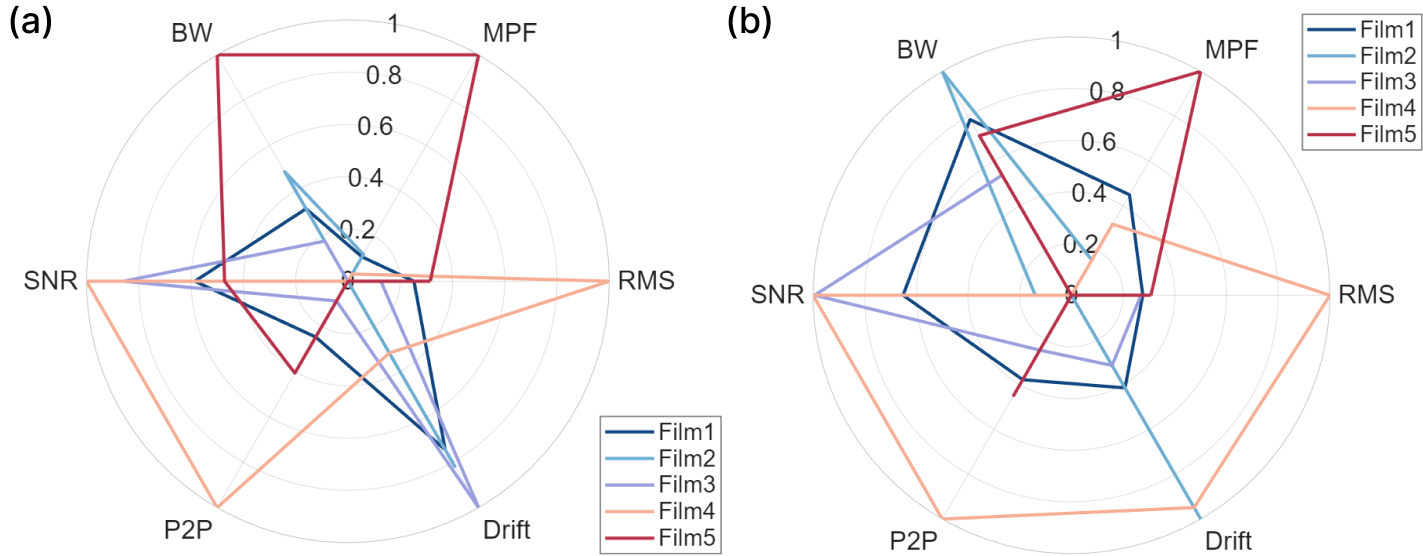}
    \caption{Feature comparison of different waterproof membranes during finger extension and wrist extension.}
    \label{membrane_radar}
\end{figure}

Among the tested materials, the 5~mil PE membrane achieved the best overall performance by maintaining strong vibration transmission, acceptable signal-to-noise ratio, and stable baseline characteristics. Therefore, it was selected as the waterproof interface for subsequent underwater MMG experiments.

\subsection{Underwater Experimental Evaluation}

To evaluate the underwater sensing capability of the proposed MMG sensor, experiments were conducted under five environmental conditions: air, water surface, 5~cm depth, 10~cm depth, and 10~cm depth with external water disturbance. The selected 5~mil PE membrane was used as the waterproof interface. The sensor was attached to the brachioradialis muscle, approximately 5~cm proximal to the elbow, and repeated finger-gripping motions were recorded at 600~Hz.

\begin{figure}[!htbp]
    \centering
    \includegraphics[width=0.9\linewidth]{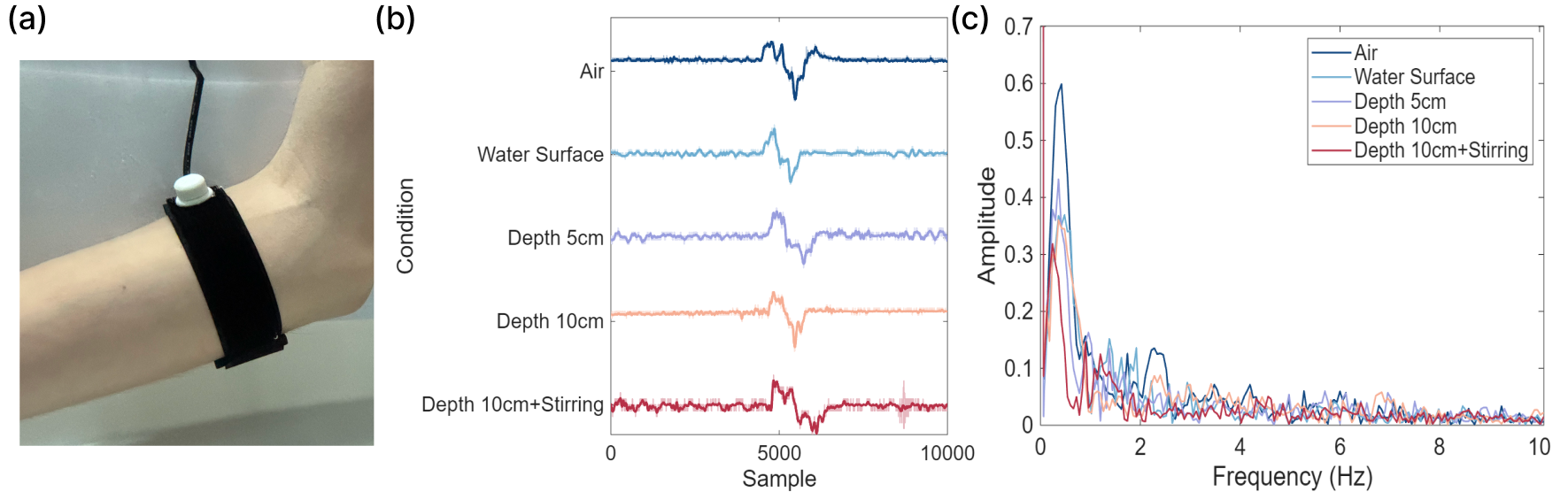}
    \caption{Underwater MMG evaluation under different environmental conditions. 
    (a) Experimental setup. 
    (b) Time-domain MMG signals under different conditions. 
    (c) Corresponding frequency responses.}
    \label{duibi}
\end{figure}

The MMG waveforms remained distinguishable across different underwater conditions. Although water immersion and external disturbance introduced additional fluctuations, contraction-related vibration patterns were preserved. The frequency responses showed consistent dominant components, indicating that the waterproof structure did not substantially alter the characteristic MMG spectrum.

\begin{figure}[!htbp]
    \centering
    \includegraphics[width=0.9\linewidth]{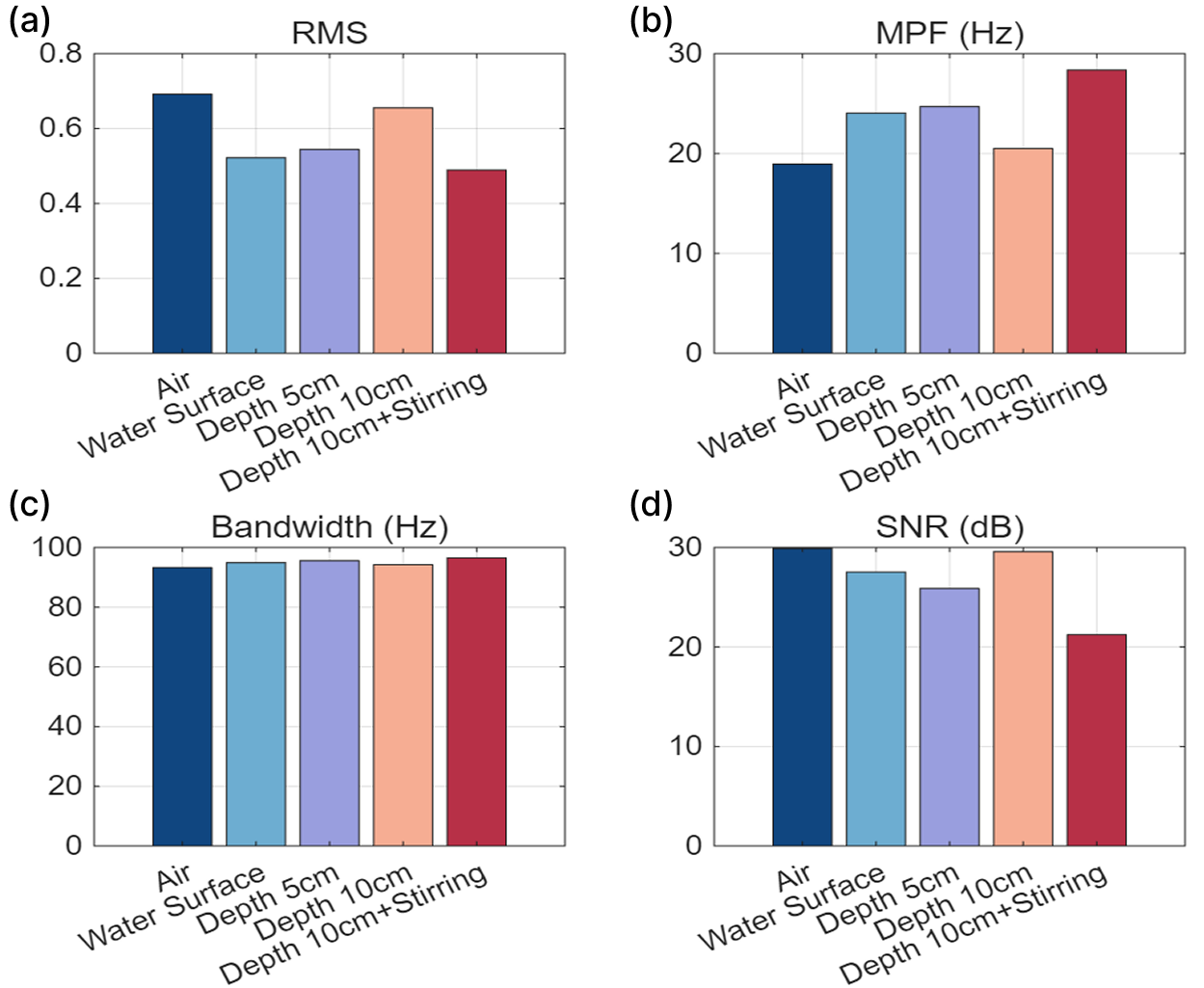}
    \caption{Quantitative comparison of underwater MMG signals. 
    (a) RMS. 
    (b) MPF. 
    (c) BW. 
    (d) SNR.}
    \label{duibi2}
\end{figure}

Four signal descriptors, including RMS, MPF, bandwidth, and SNR, were calculated to quantify signal stability. The results demonstrate that the sensor maintains comparable signal characteristics after immersion, while external disturbances mainly affect noise level rather than the muscle-induced vibration components. These findings verify the feasibility of the proposed waterproof MMG sensor for underwater muscle activity monitoring.

\begin{figure}[!htbp]
    \centering
    \includegraphics[width=0.9\linewidth]{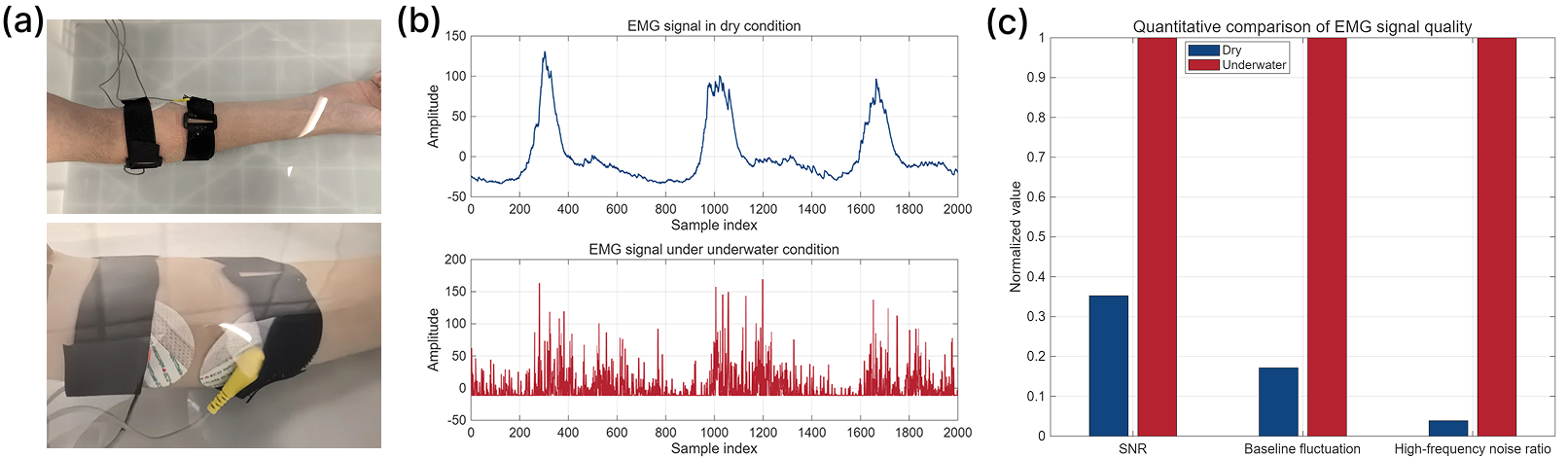}
    \caption{Comparison of EMG signal quality under dry and underwater conditions. (a) EMG measurement setup. (b) Raw EMG signals during the same muscle contraction task. (c) Signal quality comparison based on SNR, baseline fluctuation, and high-frequency noise ratio.}
    \label{emg_comparison}
\end{figure}

To further compare sensing modalities under underwater conditions, EMG measurements were conducted under dry and underwater conditions using identical contraction tasks. As shown in Fig.~\ref{emg_comparison},  underwater EMG exhibited increased fluctuations, reduced SNR, and higher high-frequency noise compared with the dry condition. These degradations are mainly attributed to unstable electrode–skin coupling and water-induced electrical disturbances, demonstrating the advantage of mechanically based MMG sensing for underwater muscle monitoring.

\subsection{Underwater Hardware System}

\begin{figure}[htbp]
\centering
\includegraphics[width=0.9\linewidth]{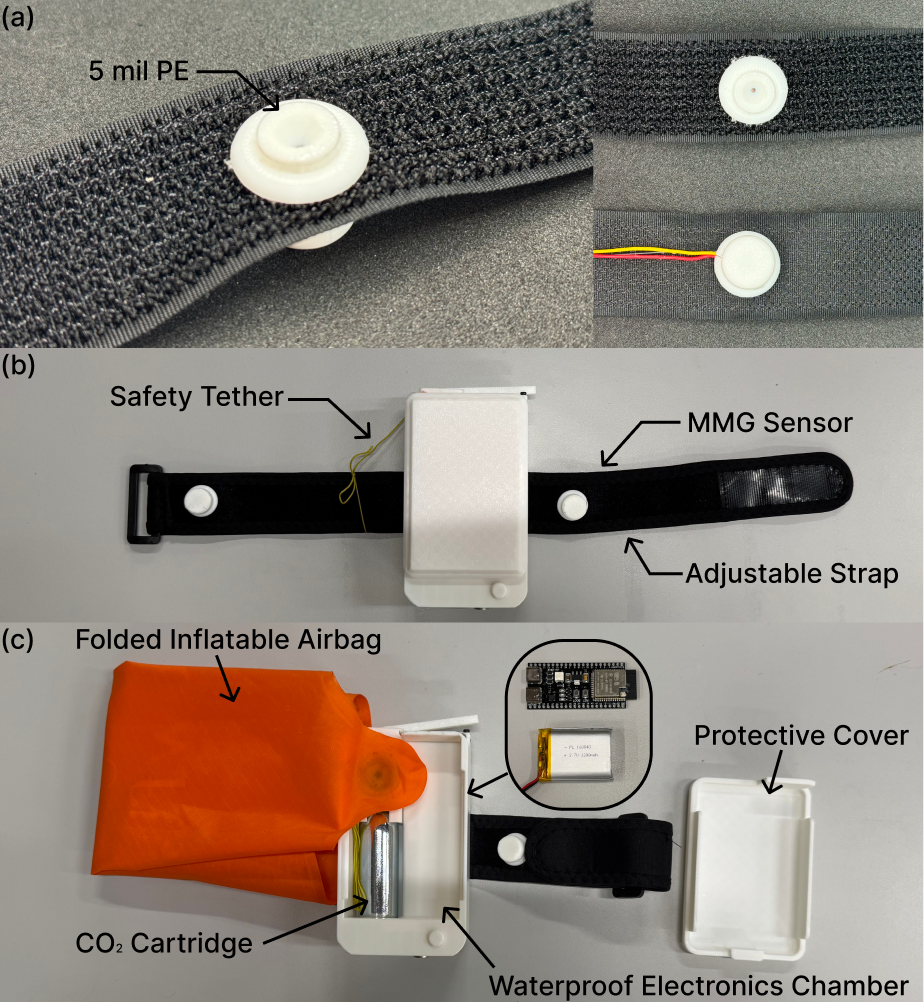}
\caption{
(a) Closed configuration of the leg-mounted device, showing the adjustable strap and waterproof electronics chamber. (b) Open configuration showing the folded inflatable airbag, CO$_2$ cartridge, waterproof chamber, ESP32-S3 control board, and battery. 
}
\label{fig:hardware_system}
\end{figure}

The underwater rescue device was designed as a leg-mounted wearable system that combines MMG-based cramp detection with automatic buoyancy deployment. As shown in Fig.~\ref{fig:hardware_system}a, the device is fixed to the lower leg by an adjustable strap, while the waterproof chamber houses the control electronics and power supply.

As shown in Fig.~\ref{fig:hardware_system}b, the buoyancy module includes a folded high-strength airbag, a standard 16 g CO$_2$ cartridge, and a solenoid-controlled release mechanism. Since the release position is located near the lower leg, the airbag is connected to the device by a lightweight safety tether, allowing it to float upward after inflation while remaining accessible to the user.

When abnormal MMG activity is detected, the ESP32-S3 activates a miniature linear solenoid lock. The solenoid controls a linkage mechanism that presses a puncturing pin into the CO$_2$ cartridge seal. The released gas inflates the airbag and opens the protective cover, providing automatic buoyancy assistance. The electronics, including the ESP32-S3, a 1200 mAh lithium battery, a TP4056 charging module, and a waterproof switch, are enclosed in a sealed 3D-printed waterproof chamber.

\section{Muscle-State Monitoring and Anomaly Detection}

\subsection{MMG Sensor Placement for Lower-Leg Cramp-Related Monitoring}
\begin{figure}[h]
\centering
\includegraphics[width=0.9\linewidth]{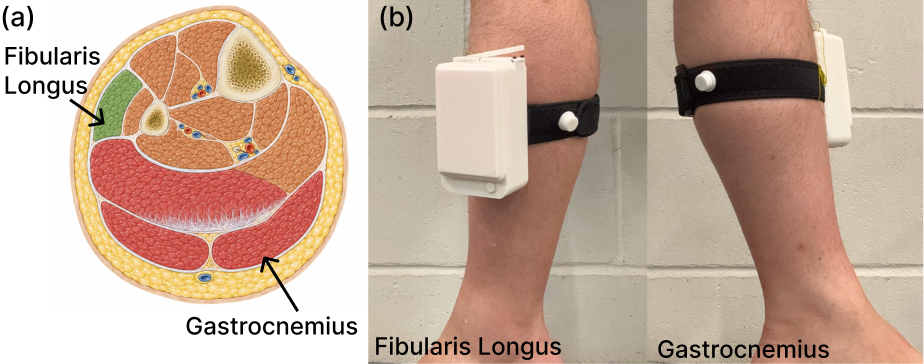}
\caption{
Sensor placement and corresponding lower-leg muscle regions. 
(a) Cross-sectional anatomical illustration highlighting the fibularis longus region and gastrocnemius region. (b) Wearable MMG sensor placement on the lateral and posterior lower-leg regions.
}
\label{place}
\end{figure}

A unilateral lower-leg configuration was adopted for practical underwater deployment, with cramp-related muscle-state monitoring as the primary objective. The lower leg was selected because swimming-related cramps commonly affect the calf region, while fatigue-related neuromuscular changes can alter lower-leg muscle activity during prolonged aquatic motion. The fibularis longus and gastrocnemius were selected as the sensing sites~\cite{Huang_Yang_2025}, as shown in Fig.~\ref{place}a. The gastrocnemius provides a superficial and cramp-prone posterior calf target, whereas the fibularis longus provides complementary lateral information related to ankle stabilisation, eversion, and plantarflexion. This pairing helps distinguish cramp-related abnormal contraction from normal rhythmic lower-leg movement.

As shown in Fig.~\ref{place}b, two MMG sensors were attached to the left lower leg, targeting the fibularis longus muscle belly and the medial gastrocnemius muscle belly, respectively. To improve mechanical signal acquisition and ensure stable waterproof attachment, both sensors were placed on soft muscle regions rather than tendon-dominant areas. The sensor wires were routed inside the strap to avoid externally exposed cables, thereby reducing cable entanglement and improving attachment stability during underwater movement.

\subsection{Representative MMG Signals Under Four Swimming Styles}

To investigate whether underwater MMG signals can reflect stroke-dependent lower-limb muscle activity, representative recordings were collected from five healthy participants, all of whom were able to perform the four common swimming styles. The demographic information of the participants is summarized in Table~\ref{tab:participants}.

\begin{table}[htbp]
\centering
\caption{Participant information.}
\label{tab:participants}
\begin{tabular}{c c c c c c}
\hline
\textbf{ID} & \textbf{Age} & \textbf{Sex} & \textbf{Weight (kg)} & \textbf{Height (cm)} & \textbf{Body type} \\
\hline
P1 & 21 & M & 76 & 175 & Slightly overweight \\
P2 & 20 & M & 72 & 178 & Normal \\
P3 & 25 & M & 64 & 176 & Lean \\
P4 & 22 & M & 73 & 179 & Normal \\
P5 & 21 & F & 58 & 162 & Lean \\
\hline
\end{tabular}
\end{table}

All measurements were conducted underwater at a depth of approximately 0.2--0.5~m. During data collection, participants maintained a floating posture while holding the pool edge or boat side for support, and an additional observer remained onshore to ensure safety.

\begin{figure}[!htbp]
    \centering
    \includegraphics[width=0.9\linewidth]{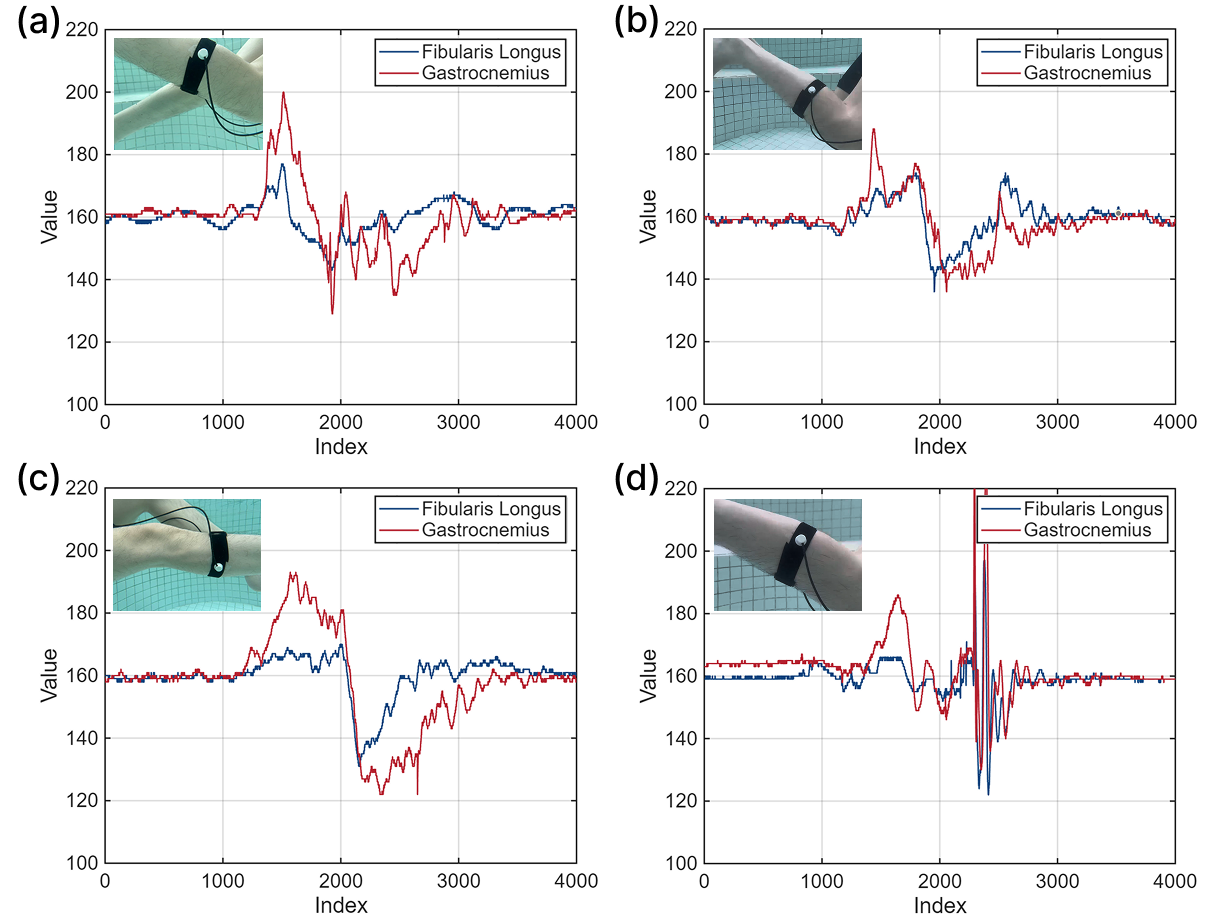}
    \caption{(a)--(d) Representative underwater MMG signals recorded from the fibularis longus and gastrocnemius during four swimming styles: (a) freestyle, (b) breaststroke, (c) backstroke, and (d) butterfly.}
    \label{swimming_mmg}
\end{figure}

The MMG signals were acquired at a sampling frequency of 1200~Hz. For each swimming style, participants performed rhythmic leg movements at an approximately constant pace of one motion cycle per second. Specifically, freestyle and backstroke involved repeated reciprocal kicking, breaststroke involved periodic frog-kick-like propulsion, and butterfly involved synchronized dolphin-like kicking.

Figure~\ref{swimming_mmg} shows representative MMG signals from the fibularis longus and gastrocnemius during freestyle, breaststroke, backstroke, and butterfly. Clear stroke-dependent differences can be observed in both amplitude and temporal pattern. Freestyle and backstroke show relatively continuous cyclic modulation, whereas breaststroke and butterfly exhibit more concentrated bursts and clearer phase transitions. In particular, butterfly produces sharper transient fluctuations, consistent with its more synchronized and explosive lower-limb action. These results suggest that underwater MMG signals preserve distinct muscle activation signatures across swimming styles, providing a basis for subsequent signal processing and motion classification.

\subsection{Representative MMG Signals During a Cramp Event}

\begin{figure}[htbp]
\centering
\includegraphics[width=0.9\linewidth]{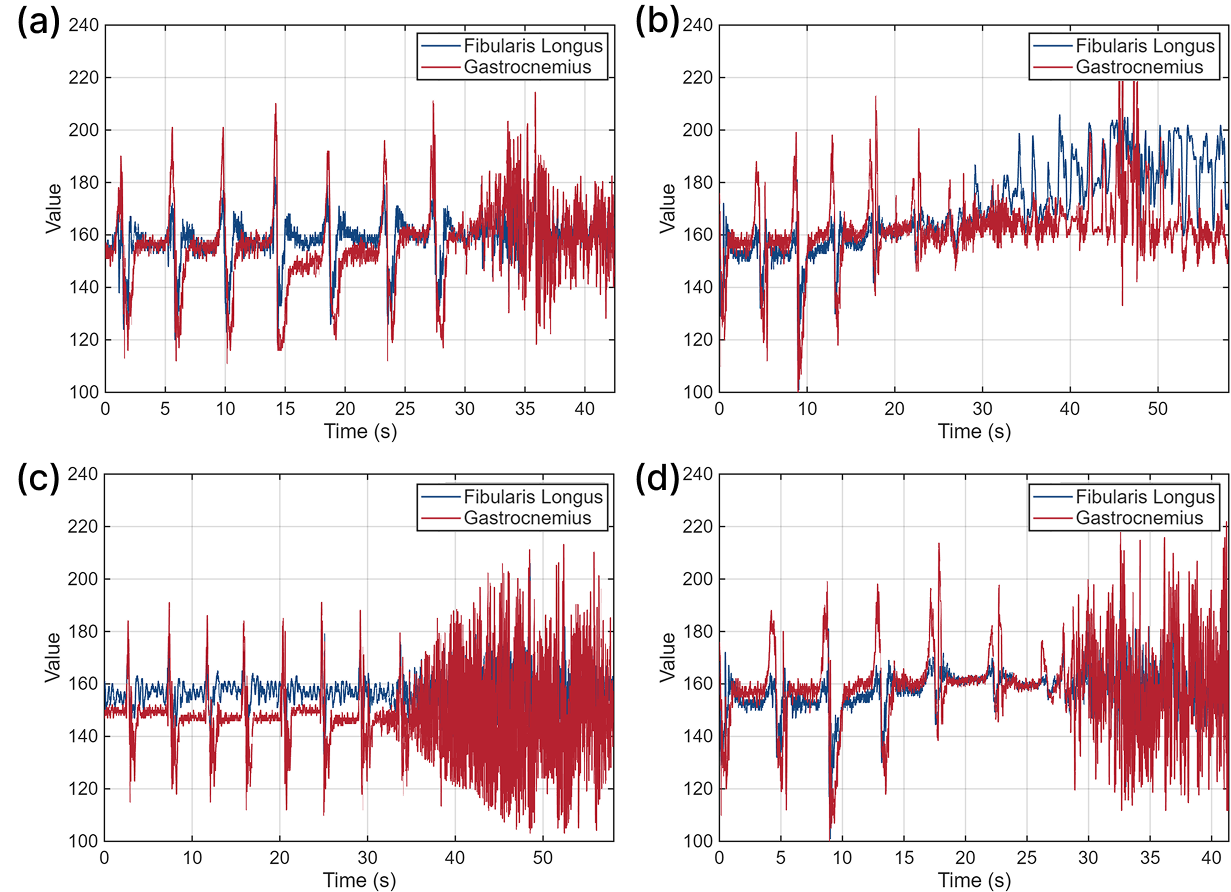}
\caption{
Representative MMG signals from four cramp-related underwater trials. 
(a)--(c) Backstroke-induced events; (d) freestyle-induced event. 
The recordings show a transition from regular rhythmic activation to irregular high-amplitude fluctuations, especially in the gastrocnemius channel, indicating abnormal involuntary contraction and altered inter-muscle coordination.
}
\label{fig:cramp_backstroke}
\end{figure}

To further evaluate whether underwater MMG can capture abnormal muscle activity beyond regular swimming patterns, four representative cramp-related underwater trials were analyzed, as shown in Fig.~\ref{fig:cramp_backstroke}. The signals were recorded from the fibularis longus and gastrocnemius muscles of the left lower leg. Among these trials, three events occurred during backstroke and one occurred during freestyle.

As shown in Fig.~\ref{fig:cramp_backstroke}, the early stage of each recording exhibits relatively regular rhythmic bursts, consistent with repeated lower-limb kicking during swimming. As the event progresses, this regular structure is disrupted, and dense irregular fluctuations appear, particularly in the gastrocnemius channel. This suggests that cramp-related abnormality is not only reflected by increased amplitude, but also by changes in waveform shape and temporal organization.

Based on this observation, three interpretable features were extracted from each segmented motion cycle: ShapeDev, WaveformLengthDev, and AmpDev. A normal-motion template $c_{ref}$ was first constructed from the initial stable cycles. For the $i$-th cycle $c_i$, ShapeDev was defined as

\begin{equation}
\mathrm{ShapeDev}_i = 1-\mathrm{corr}(c_i,c_{ref}),
\end{equation}

which measures the loss of similarity between the current cycle and the normal rhythmic pattern. WaveformLengthDev was used to describe intra-cycle irregularity,

\begin{equation}
\mathrm{WL}_i=\sum_{j=1}^{N-1}|c_i(j+1)-c_i(j)|,
\end{equation}

where a larger value indicates a more jagged and unstable waveform. AmpDev was defined by the peak-to-peak amplitude of each cycle,

\begin{equation}
A_i=\max(c_i)-\min(c_i).
\end{equation}

For comparison across trials, these features were normalized relative to the initial stable cycles.

\begin{figure}[htbp]
\centering
\includegraphics[width=0.9\linewidth]{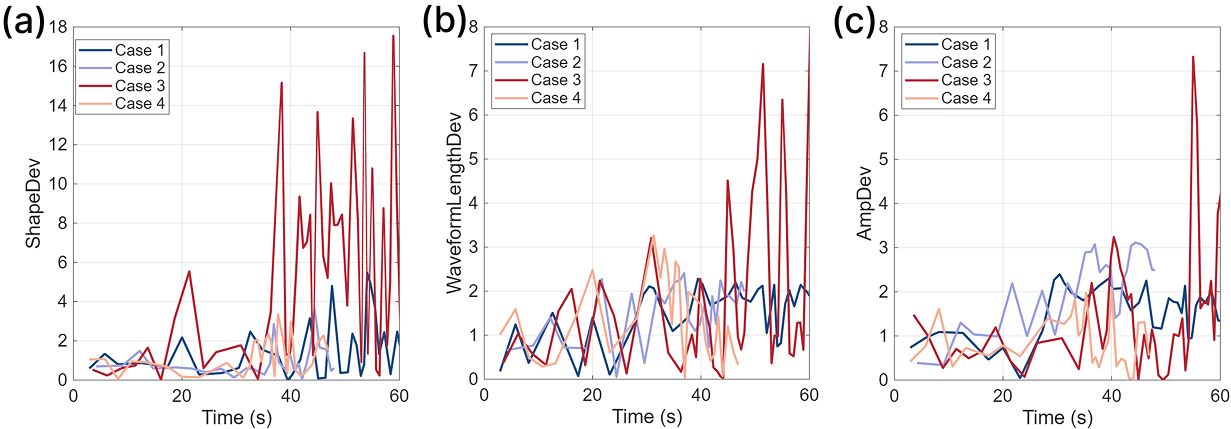}
\caption{
Feature-level analysis of the four cramp-related trials. 
(a) ShapeDev measures deviation from the normal motion template. 
(b) WaveformLengthDev captures increased waveform irregularity. 
(c) AmpDev reflects abnormal amplitude increases. 
Together, these features quantify the transition from regular swimming motion to cramp-related abnormal muscle activity.
}
\label{fig:cramp_features}
\end{figure}

As shown in Fig.~\ref{fig:cramp_features}, the three features increase during the abnormal phases of the trials, although their relative sensitivity differs across cases. This observation is consistent with the reported physiological characteristics of exercise-associated muscle cramps, which are commonly described as painful involuntary skeletal-muscle contractions during or shortly after exercise. From the altered neuromuscular control perspective, fatigue-related changes can increase motor-neuron excitability and disrupt normal muscle coordination, leading to abnormal and sustained contraction patterns. In this context, ShapeDev reflects the loss of regular swimming-cycle organization, WaveformLengthDev captures the denser and more unstable waveform fluctuations, and AmpDev reflects abrupt contraction-related amplitude increases. Therefore, the observed MMG feature changes are physiologically consistent with cramp-related abnormal muscle activity and provide a simple quantitative basis for the subsequent cramp-risk score calculation.

\section{Experimental Validation}
\label{introduction}

\subsection{Swimming Stroke Classification}

To provide reliable motion context for subsequent cramp detection, a lightweight multivariate time-series classification framework was developed to identify the current swimming stroke from two lower-leg MMG channels. The classification task considered four stroke classes: butterfly (FLY), breaststroke (BR), backstroke (BK), and freestyle (FR). The full dataset contained 275 trials, including 80 butterfly, 70 breaststroke, 58 backstroke, and 67 freestyle recordings. Each trial was treated as an independent file and split at the file level into training, validation, and test subsets with a ratio of 0.7/0.15/0.15, so as to prevent information leakage between overlapping windows extracted from the same recording.

Let the two-channel MMG signal of one trial be denoted as
\[
\mathbf{X}=\{x_c(t)\in\mathbb{R}\mid c\in\{1,2\},\; t=1,\dots,T\},
\]
where \(c\) denotes the sensor channel and \(T\) is the total number of samples in the trial. A sliding-window strategy was adopted to segment each trial into a sequence of fixed-length samples. For a window size \(L=800\) and stride \(S=400\), the \(i\)-th window is defined as
\[
\mathbf{W}_i=\mathbf{X}[:,\, (i-1)S+1:(i-1)S+L], \qquad i=1,2,\dots,N,
\]
where \(N\) is the total number of extracted windows in the trial. Using this strategy, the training, validation, and test sets contained 780, 172, and 173 windows, respectively.

Each window \(\mathbf{W}_i\in\mathbb{R}^{2\times L}\) was then fed into a MiniRocket-based classifier. In this framework, the multivariate MMG window is transformed into a high-dimensional feature representation through a set of fixed convolutional kernels,
\[
\mathbf{z}_i=\phi(\mathbf{W}_i),
\]
where \(\phi(\cdot)\) denotes the MiniRocket feature transformation. In the present implementation, the number of kernels was set to 6000, with a maximum dilation of 32. Based on the transformed feature vector \(\mathbf{z}_i\), the classifier predicts the corresponding stroke label as
\[
\hat{y}_i=\arg\max_{k\in\{1,\dots,4\}} f_k(\mathbf{z}_i),
\]
where \(f_k(\mathbf{z}_i)\) denotes the decision score for class \(k\). The random seed was fixed at 42 for reproducibility.

To obtain a robust trial-level decision, a temporal voting strategy was further applied over all windows belonging to the same file. Specifically, the final predicted label of one trial was determined by majority voting:
\[
\hat{Y}=\arg\max_{k\in\{1,\dots,4\}} \sum_{i=1}^{N}\mathbb{I}(\hat{y}_i=k),
\]
where \(\mathbb{I}(\cdot)\) is the indicator function. This procedure reduces the influence of locally ambiguous windows and yields a more stable file-level classification result.

The proposed classifier achieved a window-level accuracy of 91.86\% and a file-level accuracy of 95.12\% on the validation set. On the test set, the window-level accuracy reached 91.91\%, while the file-level accuracy further increased to 97.56\%, as summarized in Table~\ref{tab:stroke_cls}. These results indicate that although some local windows may be difficult to distinguish, the overall stroke type of a complete trial can be identified more robustly after temporal voting after temporal voting.

\begin{table}[htbp]
\centering
\caption{Performance of the MiniRocket-based swimming stroke classifier.}
\label{tab:stroke_cls}
\begin{tabular}{lcc}
\hline
Metric & Validation Set & Test Set \\
\hline
Window-level accuracy & 91.86\% & 91.91\% \\
File-level accuracy   & 95.12\% & 97.56\% \\
\hline
\end{tabular}
\end{table}

\begin{figure}[htbp]
    \centering
    \includegraphics[width=0.9\linewidth]{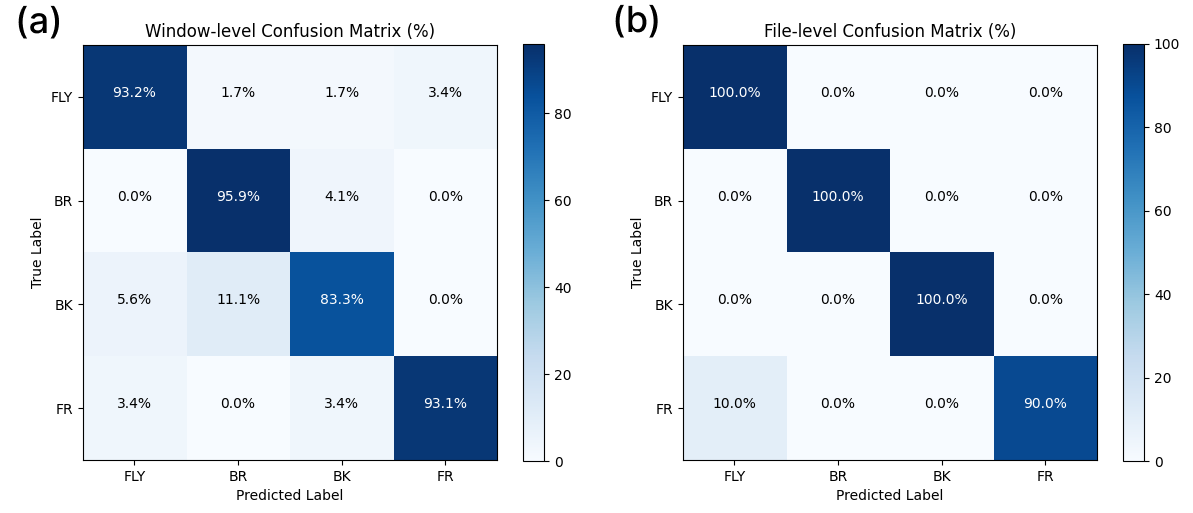}
    \caption{Confusion matrices of the MiniRocket-based swimming stroke classifier. Left: window-level results. Right: file-level results after temporal voting. Most residual confusion occurred for backstroke at the window level, while file-level voting improved robustness and yielded near-perfect trial-level classification.}
    \label{fig:stroke_confusion}
\end{figure}

As shown in Fig.~\ref{fig:stroke_confusion}, butterfly, breaststroke, and freestyle were recognized with strong consistency at the window level, while backstroke was relatively more difficult to classify. In particular, backstroke was occasionally misclassified as breaststroke (11.11\%) or butterfly (5.56\%), suggesting partial overlap in local MMG patterns across these stroke phases. Freestyle also exhibited a small confusion with butterfly (3.45\%). Nevertheless, after file-level voting, the classifier achieved perfect recognition for butterfly, breaststroke, and backstroke, while freestyle reached 90.0\% recall, with only one trial being misclassified as butterfly.

Overall, these results demonstrate that the proposed MiniRocket-based framework provides a sufficiently accurate and computationally efficient solution for online swimming-stroke recognition, thereby establishing a reliable motion-context input for the subsequent cramp-related monitoring module.

\subsection{Pattern-Based Cramp-Related Muscle-State Monitoring}

To identify cramp-related abnormal muscle-state transitions that emerge during continuous swimming rather than during static or low-activity phases, a pattern-based monitoring strategy was developed. Instead of directly thresholding signal amplitude, the proposed method first models the regular motion pattern and then identifies deviations from this learned pattern.

Let \(x_1(t)\) and \(x_2(t)\) denote the two MMG channels. After smoothing, the Hilbert envelopes of the two channels were extracted and fused as
\[
e_f(t)=\frac{1}{2}\left(\tilde{e}_1(t)+\tilde{e}_2(t)\right),
\]
where \(\tilde{e}_1(t)\) and \(\tilde{e}_2(t)\) are the normalized envelopes of the two channels. Peaks detected from \(e_f(t)\) were then used to segment the signal into successive motion cycles.

The first \(K\) normal cycles were used to establish a reference template,
\[
\mathbf{c}_{\mathrm{ref}}=\frac{1}{K}\sum_{i=1}^{K}\mathbf{c}_i,
\]
where \(\mathbf{c}_i\) denotes the normalized waveform of the \(i\)-th cycle.

For each subsequent cycle, its waveform similarity to the learned template was quantified as
\[
\rho_i=\mathrm{corr}(\mathbf{c}_i,\mathbf{c}_{\mathrm{ref}}).
\]
In addition, cycle length, amplitude, intra-cycle standard deviation, and waveform length were extracted to describe temporal and amplitude-related deviations.

These deviation terms were then combined into a cramp risk score:
\begin{equation}
\begin{aligned}
S_i=
0.30\,z_i^{(\mathrm{shape})}
+0.20\,z_i^{(\mathrm{len})}
+0.20\,z_i^{(\mathrm{amp})}\\
+0.15\,z_i^{(\mathrm{std})}
+0.15\,z_i^{(\mathrm{wl})},
\end{aligned}
\end{equation}
where \(z_i^{(\mathrm{shape})}\), \(z_i^{(\mathrm{len})}\), \(z_i^{(\mathrm{amp})}\), \(z_i^{(\mathrm{std})}\), and \(z_i^{(\mathrm{wl})}\) denote the normalized deviation scores of waveform shape, cycle length, amplitude, standard deviation, and waveform length, respectively.

The resulting score sequence was interpreted by a four-state machine: \textit{Normal}, \textit{Suspect}, \textit{Confirm}, and \textit{Deploy}. The \textit{Suspect} state indicates the emergence of abnormal muscle activity, whereas the \textit{Confirm} state requires the abnormal pattern to persist across successive cycles. Once the confirmed abnormality exceeds the deployment criterion, the system enters the \textit{Deploy} state and sends an actuation command to the emergency buoyancy module. In this way, cramp onset was identified as a pattern-breaking event within an otherwise rhythmic motion sequence, rather than as an isolated amplitude spike.

\begin{figure}[htbp]
\centering
\includegraphics[width=0.9\linewidth]{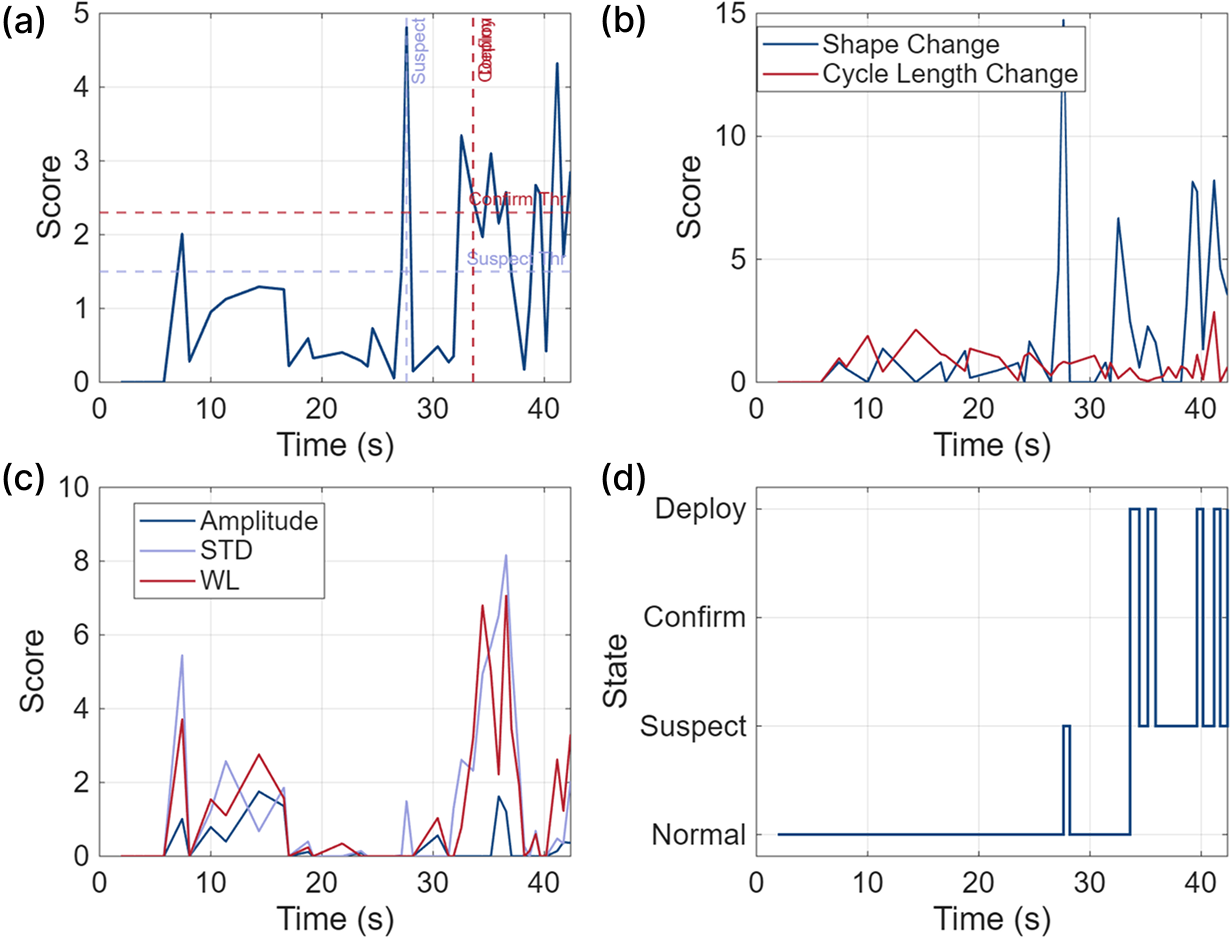}
\caption{Pattern-based monitoring result for the representative trial in Fig.~\ref{fig:cramp_backstroke}a. (a) Cramp-risk score. (b) Shape and cycle-length changes. (c) Amplitude, standard-deviation, and waveform-length changes. (d) State-machine output. The \textit{Suspect} and \textit{Confirm}/\textit{Deploy} states were reached at 27.61~s and 33.60~s, respectively.
}
\label{fig:cramp_detection}
\end{figure}

For preliminary validation, the proposed monitoring strategy was evaluated on the representative cramp-related trial shown in Fig.~\ref{fig:cramp_backstroke}a, and the corresponding state-machine response is shown in Fig.~\ref{fig:cramp_detection}. The first Suspect state appeared at 27.61~s, indicating the initial deviation from the previously learned rhythmic motion pattern. As the abnormal waveform pattern continued to develop, the state machine entered the Confirm and Deploy states at 33.60~s. This representative result suggests that the proposed strategy can identify early cramp-related muscle-state transitions during ongoing rhythmic motion and convert the detected abnormality into a deployable control decision. Overall, these preliminary results demonstrate the feasibility of combining motion-context recognition, pattern-based anomaly analysis, and state-machine decision making for in-motion cramp-related monitoring and emergency buoyancy triggering.

\subsection{System-Level Workflow for Emergency Buoyancy Triggering}
\begin{figure}[h]
\centering
\includegraphics[width=0.9\linewidth]{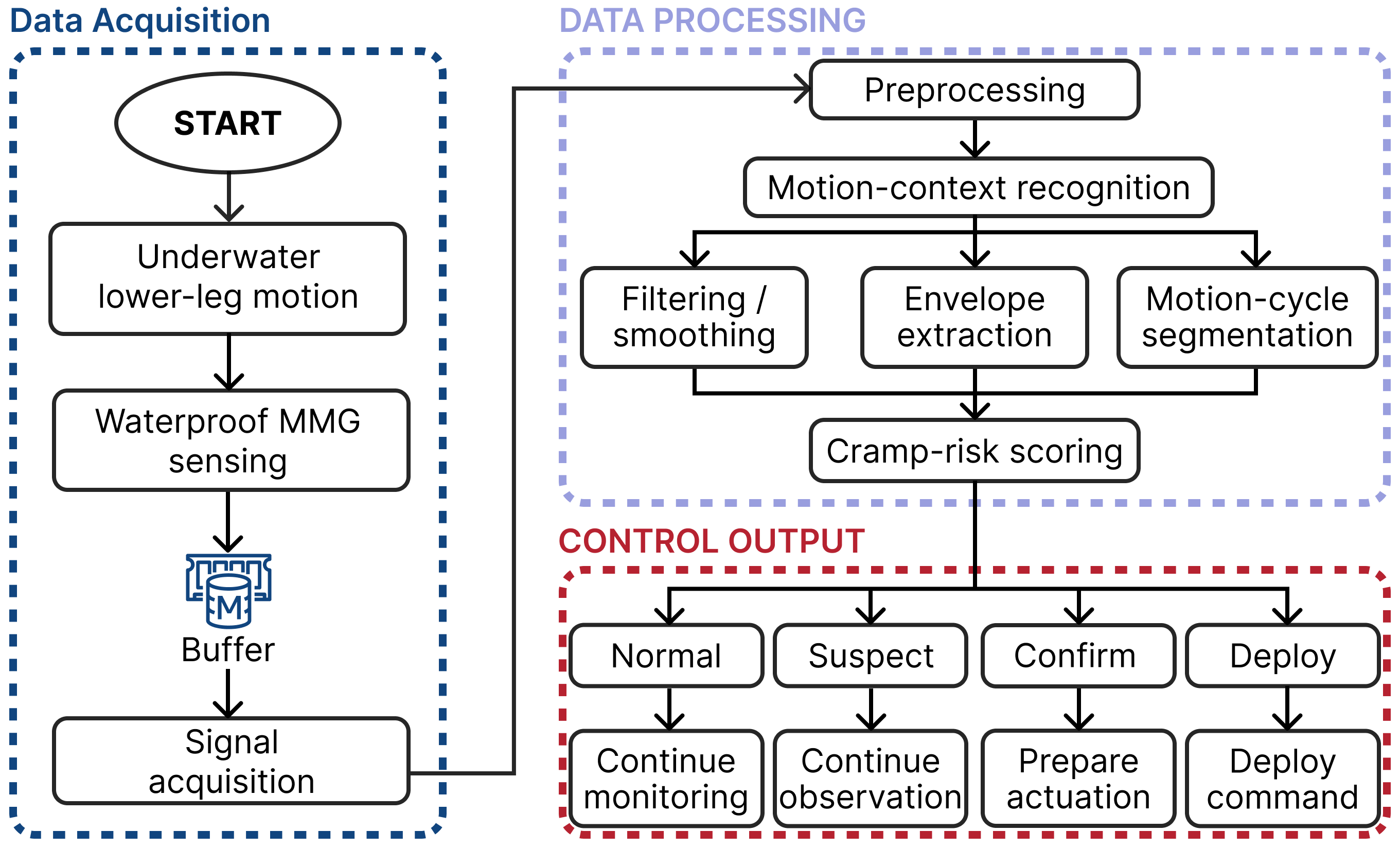}
\caption{Workflow of the underwater MMG-based emergency assistance system, from MMG acquisition and cramp-risk scoring to state-machine-triggered airbag deployment.}
\label{fig:system_workflow}
\end{figure}
The system-level workflow is shown in Fig.~\ref{fig:system_workflow}. It connects underwater MMG acquisition, motion-context recognition, cramp-risk scoring, and state-machine control into a closed sensing--decision--actuation loop.

The main purpose of this workflow is early intervention. Instead of waiting until cramp fully disrupts underwater movement, the system uses the \textit{Suspect} and \textit{Confirm} states to identify persistent pre-cramp abnormal muscle activity. Once the confirmed abnormality reaches the deployment criterion, the \textit{Deploy} state sends a command to the solenoid-controlled release mechanism. The solenoid punctures the CO$_2$ cartridge, inflates the folded airbag, and provides buoyancy assistance in less than 5~s.

\subsection{Controlled Underwater System Validation}
To evaluate the complete system under safety-constrained aquatic conditions, a controlled underwater validation test was conducted in a 60~cm-deep water tank under continuous human supervision. Only the participant's lower leg was immersed in water, allowing the sensing, decision-making, and emergency actuation functions to be evaluated while avoiding the risks associated with inducing leg cramps in deep water.

\begin{figure}[htbp]
\centering
\includegraphics[width=0.9\linewidth]{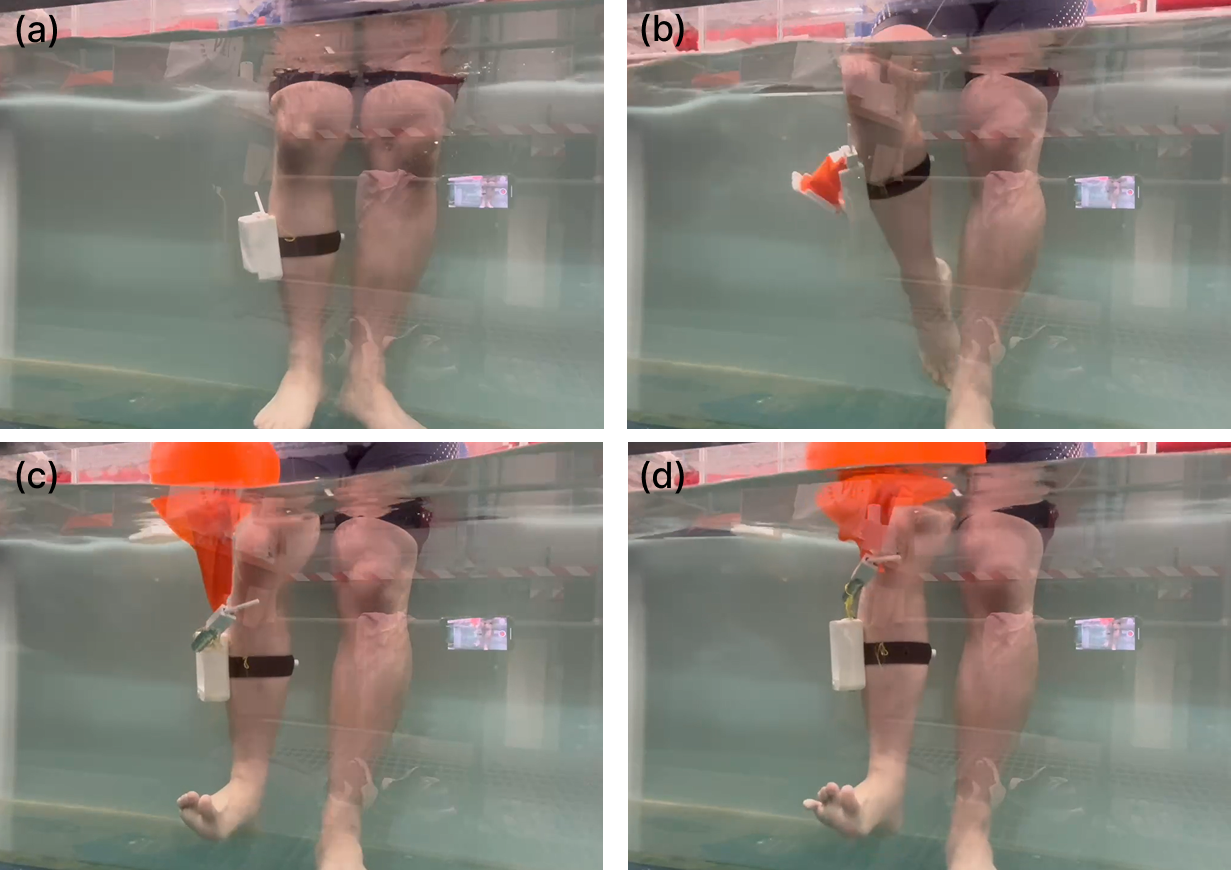}
\caption{Controlled underwater validation showing (a) initial device attachment, (b) repeated backstroke-like kicking, (c) emergency airbag deployment, and (d) inflated airbag flotation with safety tether connection.}
\label{fig:system_validation}
\end{figure}

As shown in Fig.~\ref{fig:system_validation}, the participant first performed repeated leg-kicking motions to simulate backstroke-like underwater movement. After approximately 30~min of repeated motion, gradual lower-leg stiffness and discomfort were reported, followed by involuntary twitching and numbness-like sensations. Compared with the regular kicking pattern, the MMG signals showed abnormal muscle activity, and the system triggered the emergency actuation module once the predefined criterion was reached.

After activation, the CO$_2$ release mechanism inflated the safety airbag within approximately 5~s. The airbag expanded and floated to the water surface while remaining connected to the leg-mounted device through the safety tether, making the buoyancy module visible and accessible. The deployment test was repeated five times under the same condition, and all trials successfully triggered CO$_2$ release and airbag inflation. These results demonstrate the repeatability of the sensing--decision--actuation chain under controlled aquatic exposure and suggest the potential of the system for early response to cramp-related muscle abnormalities.

\section{CONCLUSIONS}
This paper presented an underwater MMG-based wearable system for lower-leg muscle-state monitoring and emergency buoyancy assistance. A compact microphone-based MMG sensor was redesigned with a flexible waterproof membrane, and the 5 mil PE film showed a suitable balance between waterproof sealing and mechanical vibration transmission. The sensor preserved identifiable MMG responses under air, water-surface, depth-variation, and stirring conditions. With two sensors placed over the fibularis longus and gastrocnemius regions, underwater MMG signals captured stroke-dependent lower-leg activity across four swimming styles. The MiniRocket-based classifier achieved 91.91\% window-level accuracy and 97.56\% file-level accuracy on the test set, supporting the use of underwater MMG as a motion-context input for subsequent abnormality detection.

For cramp-related monitoring, representative trials showed that abnormal events were associated with a transition from regular rhythmic activation to irregular high-amplitude fluctuations, especially in the gastrocnemius channel. Based on this observation, a pattern-based cramp-risk score was developed using waveform-shape, cycle-length, amplitude, standard-deviation, and waveform-length deviations. A controlled 60 cm-deep water-tank test further validated the full system response, including underwater MMG sensing, abnormal muscle-state detection, CO$_2$ release, airbag inflation, and flotation within approximately 5 s. Although further validation with larger participant groups and more realistic aquatic conditions is still required, the results demonstrate the potential of underwater MMG for identifying early cramp-related muscle-state abnormalities and supporting wearable emergency assistance.

\addtolength{\textheight}{-12cm}   




\bibliographystyle{plain} 
\bibliography{ref}

\end{document}